%% file: epsilon-uHa.tex
\documentclass{aa}
\usepackage{txfonts}
\usepackage{natbib}
\usepackage{graphicx}

\newcommand{\zav}[1]{\left(#1\right)}
\newcommand{\hzav}[1]{\left[#1\right]}

\bibpunct[, ]{(}{)}{;}{a}{}{,}

\def\uma{$\varepsilon$~UMa}
\input{common-def}

\begin{document}

\title{Modelling the light variability of the Ap star $\varepsilon$~Ursae Majoris}

\author{D. Shulyak\inst{1} \and J. Krti{\v c}ka\inst{2}, Z. Mikul{\' a}{\v s}ek\inst{2,3}, O. Kochukhov\inst{4} \and T. L\"uftinger\inst{5}}
\offprints{D. Shulyak, \\
\email{denis.shulyak@gmail.com}}
\institute{Institute of Astrophysics, Georg-August-University, Friedrich-Hund-Platz 1, D-37077 G\"ottingen, Germany \and
Department of Theoretical Physics and Astrophysics, Masaryk University, Kotl{\' a}{\v r}sk{\' a} 2, 611 37 Brno, Czech Republic \and
Observatory and Planetarium of J. Palisa, V{\v S}B~--~Technical University, Ostrava, Czech Republic \and
Department of Physics and Astronomy, Uppsala University, Box 516, 751 20, Uppsala, Sweden \and
Institut f\"ur Astronomie, Universit\"at Wien, T\"urkenschanzstra{\ss}e 17, 1180 Wien, Austria}

\date{Received / Accepted}

\abstract
{}
{%
%In this study 
We simulate the light variability of the Ap star \object{\uma} using the observed surface distributions
of Fe, Cr, Ca, Mn, Mg, Sr, and Ti obtained with the help of the Doppler imaging technique.}
{Using all photometric data available, we specified light variations of \uma\ modulated by its rotation from far UV to IR.
We employed the \llm\ stellar model atmosphere code to predict the light variability in different photometric systems.}
{The rotational period of \uma\ is refined to $5\fd088631(18)$.
It is shown that the observed light variability can be explained as a result of the redistribution of radiative flux
from the UV spectral region to the visual caused by the inhomogeneous surface distribution of chemical elements.
Among seven mapped elements, only Fe and Cr contribute significantly  to the amplitude of the observed
light variability. In general, we find very good agreement between theory and observations. 
We confirm the important role of Fe and Cr in determining the magnitude of the well-known depression around $5200$~\AA\ 
by analyzing the peculiar $a$-parameter. Finally, we show that the abundance spots of considered elements cannot explain
the observed variabilities in near UV and $\beta$ index, which probably have other causes.}
{The inhomogeneous surface distribution of chemical elements can explain most of the observed
light variability of the A-type CP star \uma.}


\maketitle

\section{Introduction}
Chemically peculiar (CP) stars are main-sequence A- and B-type stars that have a number of characteristic
photometric and spectroscopic anomalies. Slow rotation (and, consequently, weak meridional currents),
globally structured magnetic fields, and weak convection allow
a build up of the prominent abundance anomalies and inhomogeneities as a result of particle diffusion driven by the radiation field
\citep[first proposed by][]{michaud1970}.
These inhomogeneities are frequently seen as abundance spots on stellar surfaces 
\citep[see, for example,][]{khokhlova2000,ok2004,lehmann2007}, and as clouds of chemical elements concentrated at
different heights in stellar atmospheres \citep[][and references therein]{str3,str4,vip,hd24712}.

Among the defining characteristics of CP stars, their light variability in different photometric systems still awaits 
quantitative explanation. That this variability is seen in independent systems, simultaneously in
narrow- and broad-band bands, and in anti-phase in far-UV and optical spectral regions
suggests that the global flux redistribution caused by the phase-dependent absorption 
is the probable cause of the observed phenomena \citep{molnar1973,molnar1975}. 
If so, there may be a point \citep[or points or even regions, see][]{miksao} in the spectra where the flux remains almost unchanged, and
vary more or less in anti-phase on both sides from it. This region, called a ``null wavelength region'',
has been noted and confirmed by a number of studies \citep[see, for example,][]{leckrone1974,molnar1973,jamar1977,
sokolov2000,sokolov2006}. However, that the truly ``null wavelength region''
in stars with more complex light curves may not exist \citep{sokolov2006,sokolov2010}.
The connection between
the abundance spots and flux redistribution has emerged as a natural explanation of
the rotationally modulated light curves of CP stars, but the details of this relation still have to be determined 
on a solid theoretical basis.

With the appearance of detailed abundance Doppler images (DI) of the stellar surfaces \citep[e.g.,][]{lueftinger2003,ok2004,lehmann2007}
and advanced computers it became possible to numerically explore the role of uneven distributions of chemical elements
as a source of the light variability. First, \citet{lc-hd37776} succeeded in reproducing the light variability of a hot ($\teff=22000$~K)
Bp star HD~37776 based on the DI maps of He and Si derived by \citet{khokhlova2000}. Later, \citet{lc-hr7224} successfully fitted
the light curves of a cooler ($\teff=14500$~K) Bp star HR~7224 using Fe and Si maps presented by \citet{lehmann2007}. The main result of both
studies is a very good, parameter-free theoretical explanation of both the shapes and amplitudes of the observed light curves.
The authors thus proved, on the one side, that an inhomogeneous surface distribution of chemical elements accounts for most of the light variability
in these stars, and, on the other side, they independently confirmed the results of DI mapping.
In spite of significant progress in this direction, the role of abundance spots (and individual elements) in CP stars cooler than B-type
remains unknown.

In this paper, we continue to present results of the light-curve modelling of CP stars, concentrating on the brightest A-type CP star \uma\ (HD~112185, HR~4905).
This star has been studied extensively in the past and many good photometric observations are available, including the data obtained by space photometry experiments.
The surface distributions of seven elements (Ca, Cr, Fe, Mg, Mn, Sr, Ti) were derived via the DI technique by \citet{lueftinger2003}, providing the necessary basis for
theoretical modelling of the photometric variability.

\section{Observations}

The CP star \uma\ is a well-known spectral, magnetic,
and photometric variable star. The $\ion{Ca}{ii}$ K line and profuse
lines of $\ion{Cr}{ii}$ vary oppositely in strength with a period of
$5\fd0877$ \citep{guth1931,guth1934,struve1943,swen1944,deutsch1947}.
Photometric variations are observed to have the same period, but
to have double waves \citep[][etc.]{provin1953,musielok1980,pyper} with
the maximum light occurring at $\ion{Ca}{ii}$ minimum chosen to be the
phase 0.00. Far ultraviolet OAO-2 spectrometer observations of \uma\
indicate strong light variations in the anti-phase with optical ones
\citep{molnar1975,mallama1977}.

Spectropolarimetry of \uma\ has detected its very weak
variable magnetic field whose maximum coincides with the optical
light maximum \citep{borra1980,bohlender1990,donati1990,wade2000}.

Until now, almost all papers dealing with variations in the light of
\uma\ have used the \citet{guth1931} ephemeris
\begin{equation}
\mathit{HJD}(\ion{Ca}{ii}\
\mathrm{min.\,int.})=2\,426\,437.01+5\fd0887\times E,
\label{eq:guth}
\end{equation}
where $E$ is the epoch.

In this paper, we use light curves observed in different
photometric systems. Str\"omgrem $\mathit{uvby}$
measurements were taken from \citet{pyper}, and extended by
10-color medium-band Shemakha photometry {with
filters similar to standard $\mathit{uvby}$ filters
\citep{musielok1980}. $B$ and $V$ observations close to
the Johnson standard were performed by \citet{provin1953}. Unfortunately,
no reliable data has been obtained in the broad-band
$U$ filter, the available $\mathit{UBV}$ observations of
\citet{srivastava1989} been ususable due to their excessive scatter. The
same concerns available data in the $B_{\rm T}$ and $V_{\rm T}$ passbands from
Tycho catalog, while Hipparcos $H_{\rm P}$ photometric
measurements \citep{hip-tycho} are well calibrated and were used here
in the light curve modelling.

We also used  photometric data derived by
\citet{molnar1975} and \citet{mallama1977} using spectrometry of
\uma\ performed by the ultraviolet satellite OAO-2 \citep[more about the
project is given in][]{code1970}. We also attempted to use spectroscopic
observations obtained by the International Ultraviolet Explorer (IUE)
and Copernicus space missions available via the Multimission Archive at
STScl\footnote{{\tt http://archive.stsci.edu/}}. Finally, we used a
dataset of photometric observations carried out by the Wide Field
Infrared Explorer (WIRE) obtained during June-July 2000,
although we had to rejected several parts
strongly affected by instrumental effects and instabilities
\citep[see][for details]{wire}.

\section{Methods}
\label{sec:methods}

\subsection{Model atmospheres}
To perform the model atmosphere calculations,
we used the most recent version of the \llm\, \citep{llm} stellar model atmosphere code. 
For all calculations, local 
thermodynamical equilibrium (LTE) and plane-parallel geometry were assumed.
The \vald\ database \citep{vald1,vald2} was used as a main source of the atomic line data 
for computation of the line absorption coefficient. The VALD compilation contains
information about $66\times10^6$ atomic transitions. Most of them come from the latest 
theoretical calculations performed by R.~Kurucz\footnote{{\tt http://kurucz.harvard.edu}}.

The global magnetic field of the star is very weak  (of polar intensity on the order of a few hundred Gauss
\citep{donati1990,bohlender1990}), which
allows us to ignore its possible effect on the variability of the outgoing flux.

\subsection{Calculation of the light curve}

In previous attempts to simulate the light curves of HD~37776 \citep{lc-hd37776} and HR~7224 \citep{lc-hr7224},
the following approach has been taken:
\begin{enumerate}
\item
The construction of a grid of model atmospheres for a number of abundances combinations
that cover the parameter space provided by DI maps.
\item
Interpolation of specific intensities (or fluxes)
from the grid onto a combination of abundances for all meshes visible at a given phase of rotation.
The resulting flux is then obtained by surface integration of specific intensities
(or fluxes with assumed limb-darkening law).
\end{enumerate}
This procedure is applied to every rotational phase and wavelength interval occupied by a given
photometric filter. Although simple, this approach becomes challenging when modelling the light curves of \uma. 
Indeed, in both cases of HD~37776 and HR~7224, only two elements were mapped (He and Si for the former; Fe and Si for the latter).
To produce a smooth light curve, it was enough to compute a grid containing tens of models with different sets of abundances. 
Obviously, with the increasing number of mapped elements, the number of models needed to perform
abundance interpolation can become comparable or even larger than the number of surface elements originally 
used in DI calculations.

Seven mapped elements of \uma\ would require time-consuming computations.
We instead use a slightly different approach. To reduce computational expenses, we decrease
the resolution of DI maps by interpolating them from the original grid of longitudes and latitudes onto a more sparse one.
Then, for every pair of new latitude and longitude, we compute a model atmosphere with individual abundances 
that represent a given surface mesh. Next, for every model (i.e. surface mesh), we compute the light curves in a particular photometric filter
with the help of modified computer codes taken from \citet{a9-2} and extended by 
Hipparcos photometry with passbands from \citet{hip}.
The expected magnitude at a given phase $c_{\rm phase}$ is obtained by surface integration of individual fluxes $F_{i}$
from all visible meshes
\begin{equation}
c_{\rm phase} = -2.5\log\left[\displaystyle\frac{1}{U} \displaystyle \int\limits_{\mathrm{\scriptstyle visible \; surface}} F_{i} \cos\theta_{i} u(\theta_{i}) dS\right],
\label{eq:color}
\end{equation}
where $\theta$ is the angle between the normal to the surface element and the line of sight, $u(\theta)$ describes the
limb darkening for a given filter, which we assume to be a quadratic law, i.e.,
\begin{equation}
u(\theta) = 1 - a(1-\cos\theta) - b(1-\cos\theta)^2,
\label{eq:limb}
\end{equation}
and $U$ is given by
\begin{equation}
U = \displaystyle \int\limits_{\mathrm{\scriptstyle visible \; surface}}  u(\theta_i) \cos\theta_{i} dS.
\label{eq:limb2}
\end{equation}
We adopted the same limb darkening coefficients regardless of the true surface elements. In particular, the quadratic coefficients
$a$ and $b$ in each color were taken from the tables of \citet{claret2000} for the solar abundance model.
For the Hipparcos $H_{\rm P}$ filter, these coefficients were taken to be the average between the Johnson $B$ and $V$ filters.
Since there is no definite filter curve for the WIRE star tracker,
limb darkening coefficients for Johnson $V$ filter were used instead (H. Bruntt, private communication). 
Because of this uncertainty, the WIRE photometry is presented in this paper mainly for illustrative purposes.

The original resolution of DI maps of \uma\ is $33$ latitude and $68$ longitude points equidistantly spaced between 
$[0,180]$ and $[0,360]$ spherical angles. To reduce the number of calculations, we decreased the resolution of maps
to $15$ and $30$ latitude and longitude points, respectively. As proven by numerical
tests, this decrease in resolution provides abundance maps smooth enough for an accurate representation of the light curves:
we found \textit{no} visible differences in light curves when computing all $2244$ models at the original resolution, 
and $450$ models from the reduced grid (see Sect. discussion for more details).
Thus, in the remaining light curve modelling presented in this work we always used a lower number of $450$ surface elements
than the $2244$ original.

Table~\ref{tab:stellar} summarizes the basic stellar parameters adopted in the present study taken from \citet{lueftinger2003}.
The abundances are given relative to the total number ot atoms, i.e. $\varepsilon_{\rm el}=log(N_{\rm el}/N_{\rm total})$.

\input{tables/atmospheric-parameters.tex}

\section{Results}
\label{sec:results}

\subsection{Improved ephemeris}
\label{subsec:ephemeris}

\begin{figure}
\centering \resizebox{\hsize}{!}{\includegraphics{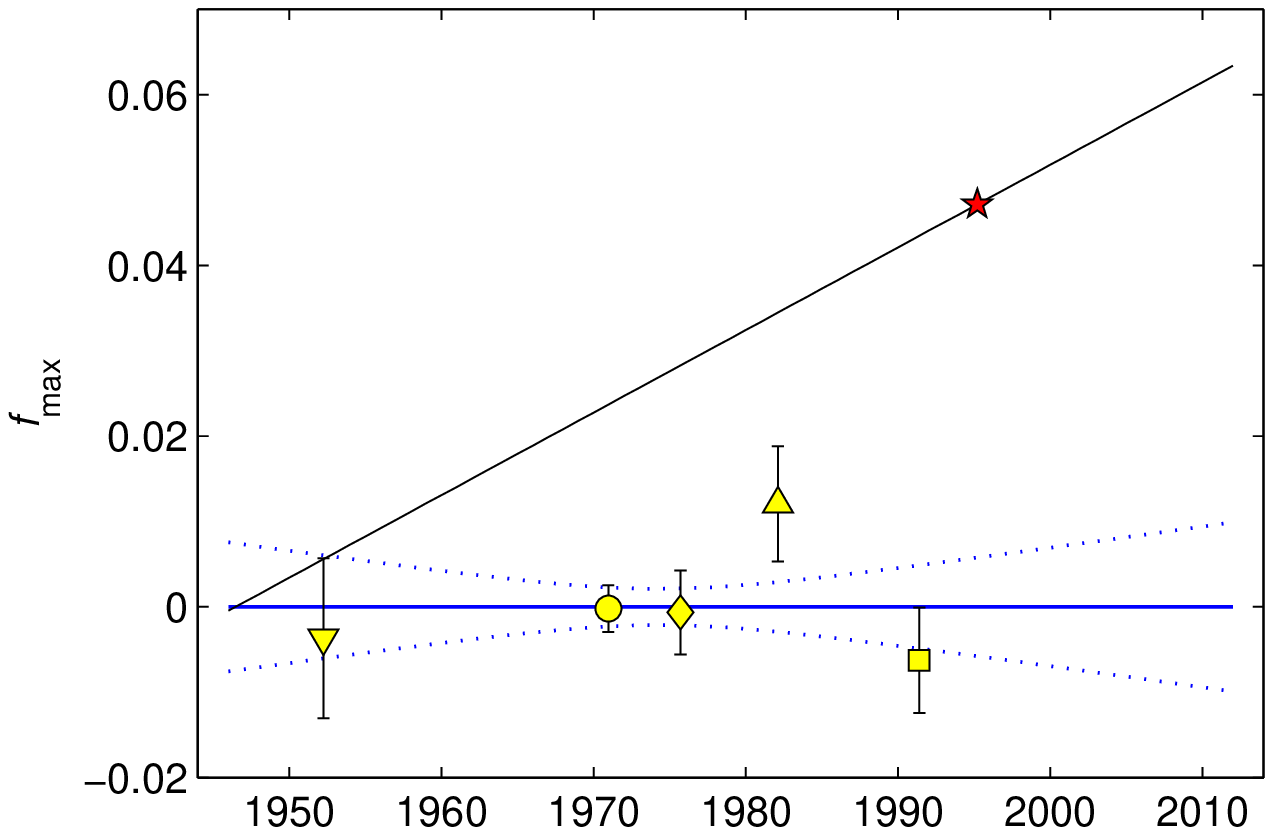}}
\caption{The dependence of the primary maximum phases derived from
Eq.\,\ref{eq:ephem} on time. $\nabla$~--~\emph{BV} photometry
\citep[][]{provin1953}, full circle -- UV photometry from OAO-2
satellite \citep[][]{molnar1975}, $\Diamond$ -- 10-color medium-band
photometry \citep[][]{musielok1988}, $\Box$ -- Hipparcos photometry
\citep[][]{hip-tycho}, $\triangle$ \emph{ubvy} photometry of
\citet{pyper}, the oblique line represents the Guthnick's ephemeris from
Eq.~(\ref{eq:guth}), and the star on it the phase position to which are
referred the spectroscopic maps we used.}\label{fig:OC}
\end{figure}

\begin{figure}
\centering \resizebox{\hsize}{!}{\includegraphics{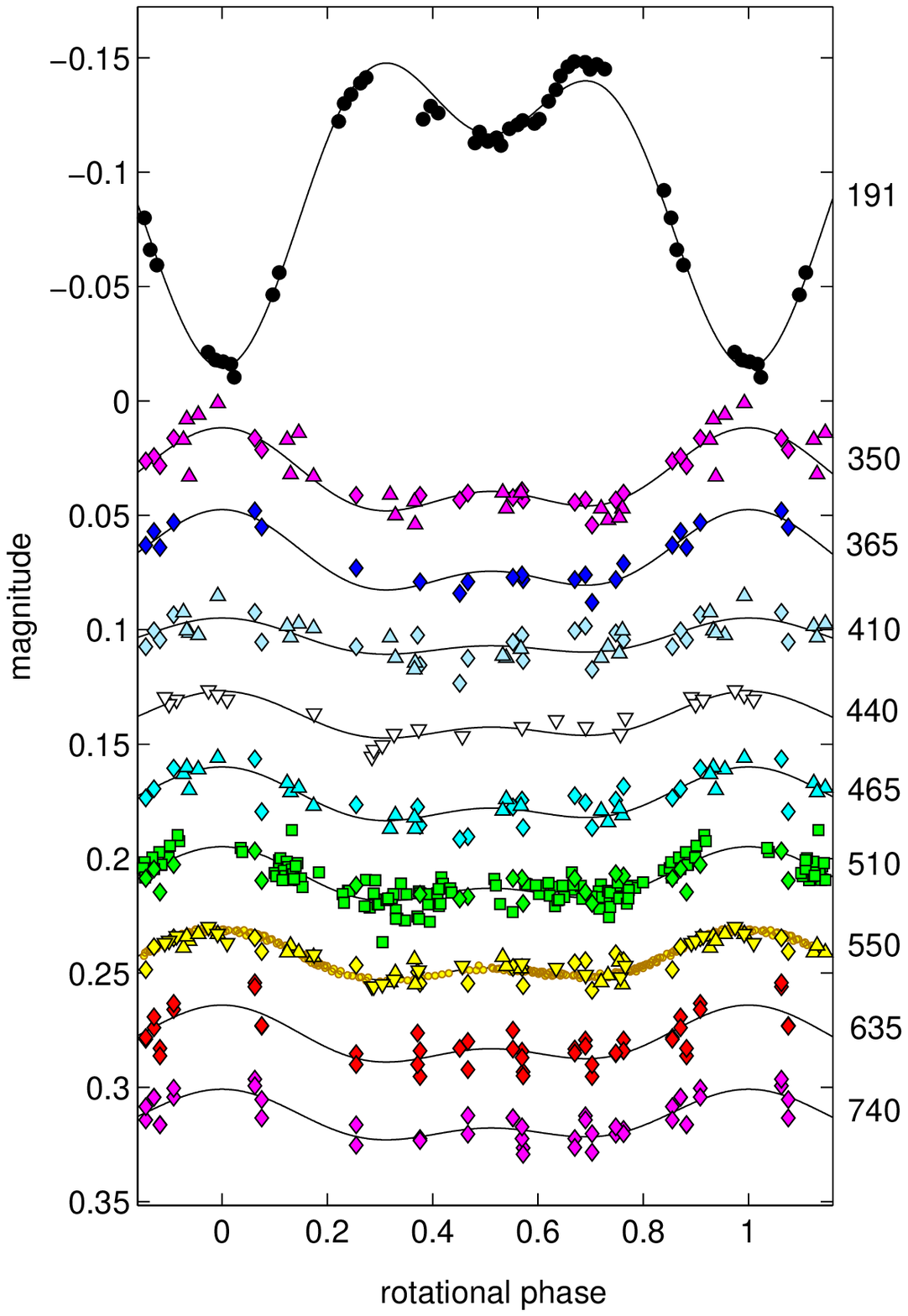}}
\caption{Observed light curves constructed from original data in
several passdands, whose effective wavelengths in nm are given on the
right. For clarity, individual LCs are shifted in their magnitudes.
Full circles -- UV photometry from OAO-2 satellite
\citep{molnar1975}, $\Diamond$~--~10-color medium-band photometry
\citep{musielok1980}, $\triangle$ \emph{ubvy} photometry of
\citet{pyper}, $\nabla\ -\ \mathit{BV}$ \citep{provin1953},
$\Box\ -$ Hipparcos photometry \citep{hip-tycho}, and small full
circles relate to broad-band photometry of WIRE \citep{wire} - each
point is the median of 1600 individual time-consecutive
measurements.} \label{fig:LC}
\end{figure}

\input{tables/phases}

Observational data used in the following analyzes of
rotationally modulated variations of \uma\ were obtained in the
past 70 years during which the star has revolved more than five
thousand times. Although we do not know the true uncertainty in the
determination of the canonical rotational period $P=5\fd0887$
published by \citet{guth1931} it can hardly be better than $\pm
0\fd0001$! This uncertainty manifests itself in the uncertainty of
$0\fp1$ in the rotational phase determination, which is for reliable
studies unacceptable.

This compelled us to improve the \uma\ rotational period.
For this purpose, we used all available reliable photometric data,
especially the 36 $BV$ measurements of \citet{provin1953}, 38 synthetic
magnitudes derived from spectrometric data in 191 nm by
\citet{molnar1973}, 179 individual observations in the ten-color
Shemakha medium-band system taken by \citet{musielok1980}, 80
Str\"omgreen $\mathit{uvby}$ measurements obtained by \citet{pyper}, andthe  314
$B_{\mathrm T}V_{\mathrm T}$, and 122 $H_{\rm P}$ measurements derived from data
of the Hipparcos satellite \citep{hip-tycho}. Thus, we had 769 individual
measurements covering the period of 1952--1993 (see Table~\ref{tab:OC}); unfortunately
measurements from the present time were absent. Measurements taken in
various narrow- and broad-band filters were divided into 10 groups according to their
effective wavelengths - see Table\,\ref{tab:ampl}.

We also used a very extensive data-set of WIRE photometry
\citep{wire}, which was unfortunately unreliable in several ways:
\begin{enumerate}
\item
We have been unable to interconnect the timing of its measurements
with standard HJD timing.
\item
Because of their large scatter and instabilities, we omitted the first part
of measurements and then one day of measurements.
\item
We found that consecutive measurements are not independent.
We therefore aggregated WIRE measurements into groups of about 1600
members. Altogether we used 195 normal points with standard
uncertainty of 0.55 mmag.
\item
We identified significant trends in the observations, but which
could be represented well by cubic polynomials and removed.
\end{enumerate}

The double-waved light curves are more or less
similar (see Fig.\,\ref{fig:LC}), differing only in their effective
amplitudes $A_c$ \citep[for a definition, see][]{mikAN}, where the
subscript $c$ denotes the photometric passband. Light-curve magnitudes can
then be expressed as
\begin{equation}\label{krivka}
 m_{cj}(t)\simeq\overline{m}_{cj}+\textstyle{\frac{1}{2}}\,A_c
 F(\vartheta),
\end{equation}
where $m_{cj}(t)$ is the magnitude in color $c$ observed by the
$j$-th observer, $\overline{m}_{cj}$ is the mean magnitude, which
can be variable over the long term. This was the case for WIRE
measurements, where we had to assume a cubic trend.

\input{tables/aeff}

\begin{figure}
\centering \resizebox{\hsize}{!}{\includegraphics{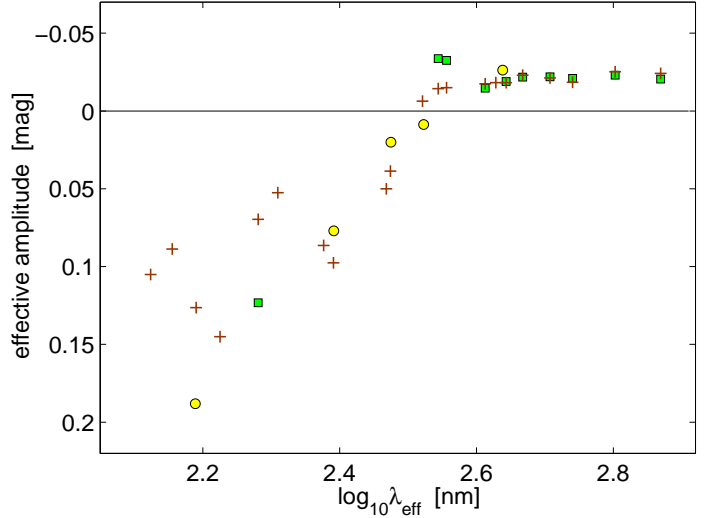}}
\caption{Effective amplitudes of observed light curves taken
in various effective wavelengths. Squares correspond to the amplitudes of LCs
taken in individual photometric filters, full
circles are values adopted from \citet{mallama1977}, and crosses are
effective amplitudes of LCs modelled in this work.}
\label{fig:obs.ampl}
\end{figure}

Function $F(\vartheta)$ is the simplest normalised periodic
function that represents the observed photometric variations of
\uma\ in detail. The phase of maximum brightness is defined to be
0.00, and the effective amplitude is defined to be 1.0. The
function, being the sum of three terms, is described by two
dimensionless parameters $\beta_1$\,and $\beta_2$, where $\beta_1$
quantifies the difference in heights of the primary and the
secondary maxima and $\beta_2$ expresses any asymmetry in the light
curve
\begin{eqnarray}\label{fce}
F(\vartheta,\, \beta_1,\,\beta_2)=
\sqrt{1\!-\!\beta_1^2\!-\!\beta_2^2}\ \cos(2\,\pi\,\vartheta)+\,
\beta_1\cos(4\,\pi\,\vartheta)+\nonumber \\
\beta_2\hzav{\textstyle{\frac{2}{\sqrt{5}}}\,\sin(2\,\pi\,\vartheta)
-\textstyle{\frac{1}{\sqrt{5}}}\sin(4\,\pi\, \vartheta)},
\end{eqnarray}
where $\vartheta$ is the phase function. The O-C diagram (see Fig.
\ref{fig:OC} and Table~\ref{tab:OC}) indicates that the function is linear; thus, we
assumed it to have the form
\begin{equation}
\vartheta=\zav{t-M_{0}}/P,\quad \mathit{HJD}(\mathrm{max\,I})=M_{0}+P\times E,
\label{eq:ephem}
\end{equation}
where $P$ is the period of the linear fit and $M_{0}$ is the HJD time of the
primary maximum nearest the weighted centre of all observations except WIRE
ones. The time of basic primary maximum in WIRE timing is then
$M_{0\rm{W}}$. All 39 model parameters were computed simultaneously by a
weighted non-linear LSM regression applied to the complete observational
material.

We found these parameters to be $P=5\fd088631(18),\ M_{0}=2\,442\,150.778(11), \mathrm{and}\ M_{0\rm{W}}=196\fd9647(31)$.
The effective amplitudes $A_c$ are given in Table\,\ref{tab:ampl},
$\beta_1=0.557(5)$, and $\beta_2=-0.056(7)$. The adequacy of the adopted
linear model for the phase function given in Eq.~(\ref{eq:ephem}) can be tested
by studying the changes in the mutual phase LC maxima with time of observed
light curves (see Fig.\,\ref{fig:OC}).

Rotational phases evaluated according to Guthnick's
ephemeris Eq.\,(\ref{eq:guth}), $\varphi_{\rm{Guth}}$ can be
transformed to new phases according to Eq.\,(\ref{eq:ephem}),
$\varphi_{\rm{new}}$ by means of the relation
\begin{eqnarray}
\varphi_{\rm{new}}&=\varphi_{\rm{Guth}}+2.648\times10^{-6}\,(t-2\,431\,960) \nonumber \\*
&=\varphi_{\rm{Guth}}+9.670\times10^{-4}\,(T-1946.4),\
\label{eq:trafo}
\end{eqnarray}
where $t$ denotes the time in JD, $T$ the time in years, and
their respectivefractions.

{Figure\,\ref{fig:obs.ampl} represents the dependence of
effective amplitudes of observed light curves on their effective
wavelengths. Values summarized in Table\,\ref{tab:ampl} were calculated using
UV effective amplitudes derived by \citet{mallama1977} from
spectrometric observations of the OAO-2 satellite. This figure clearly
illustrate the dependence of the effective amplitude on wavelength. \citet{mallama1977}
predicted that the "zero point'' should occur at a wavelength of
345\,nm. However, the "zero point'' (if any) appears to occur at a slightly shorter wavelength.

{Applying weighted LSM analysis to 28 measurements of the
effective magnetic field published in
\citet{borra1980}, \citet{bohlender1990}, \citet{donati1990}, and \citet{wade2000}, we concluded
that the observed sinusoidal variations can be satisfactorily
reproduced by the simple model of centered magnetic dipole inclined
by the angle $\beta$ to the rotational axis (see Fig.\,\ref{fig:magnet}). 
The north magnetic pole then passes the
central meridian at the phase $\varphi=-0.002(21)$, and the amplitude of
the magnetic field variations is 142(17)\,G, their mean value being
24(6)\,G. We found the relation between angles $\beta$ and the
inclination of the rotational axis with respect to the observer $i$,
to be $\tan(\beta)=3.0(9)\,\tan(i)$. Assuming \citep[together
with][]{lueftinger2003} that $i=45^{\circ}$, we found that
$\beta=72^{\circ}(6^{\circ})$, thus the north magnetic pole
is located close to the center of the leading photometric spot on \uma,
which is unlikely to be a coincidence.

The magnetic field of \uma\ is the weakest measured magnetic field in 
magnetic CP stars which would be hardly detectable in any more rapidly rotating and fainter star.
Its maximum (polar) surface magnetic field is weaker than
$400$\,Gauss. Thus we could neglect its influence in our computation.

\begin{figure}
\centering \resizebox{0.9\hsize}{!}{\includegraphics{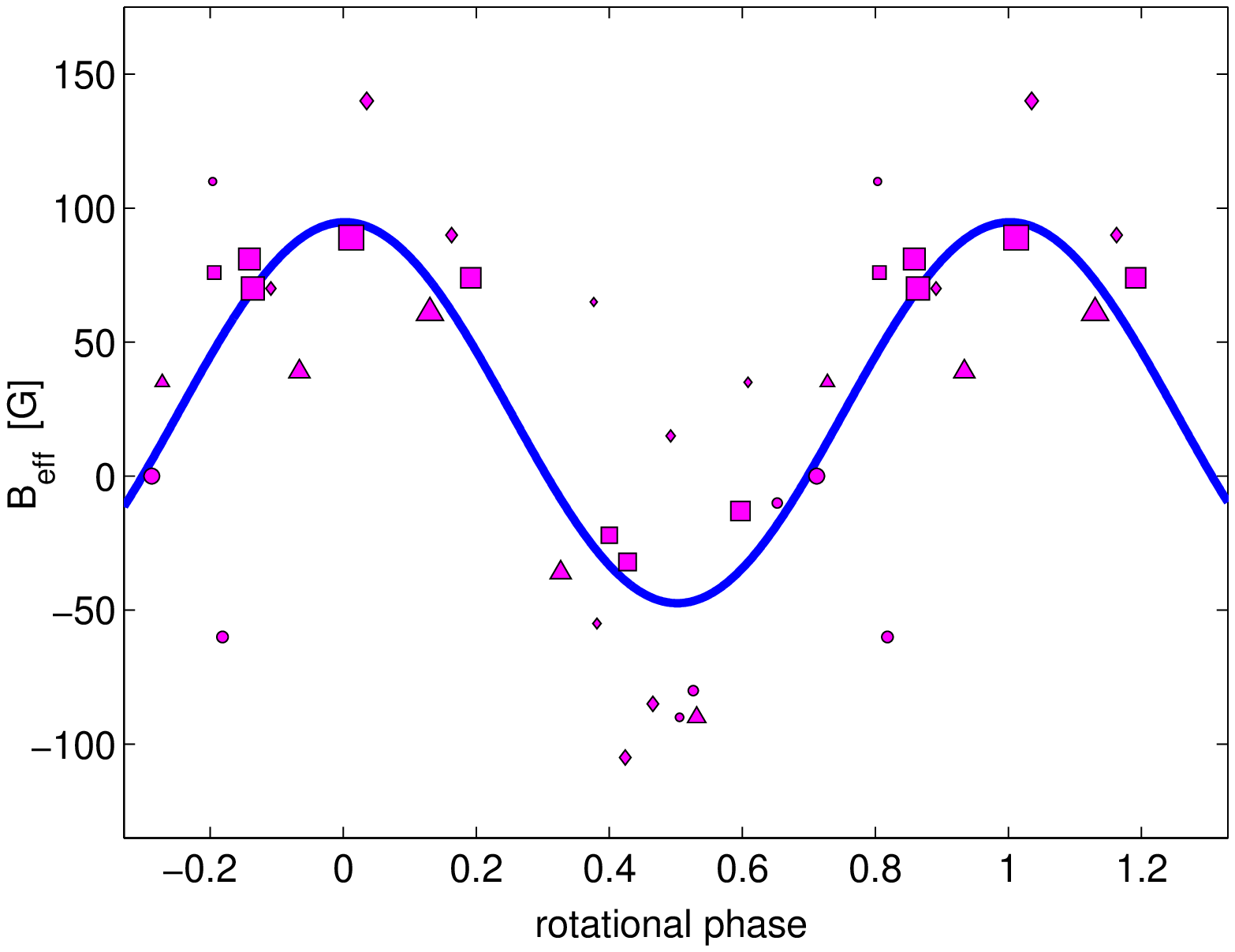}}
\caption{Effective magnetic field and its uncertainty in gauss
versus rotational phase. Note that the field is extraordinarily weak
and the phase of its maximum  coincides exactly with the maximum of
brightness. Measurements taken form: $\circ$~--~\citet{borra1980},
$\Diamond$~--~\citet{bohlender1990}, $\triangle$~--~\citet{donati1990},
$\Box$~--~\citet{wade2000}. 
Areas of individual markers correspond to the weight of the particular measurement,
which is inversely proportional to the squares of their uncertainties.}
\label{fig:magnet}
\end{figure}

\subsection{Abundance anomalies and emergent flux}
\begin{figure}
\includegraphics[width=\hsize]{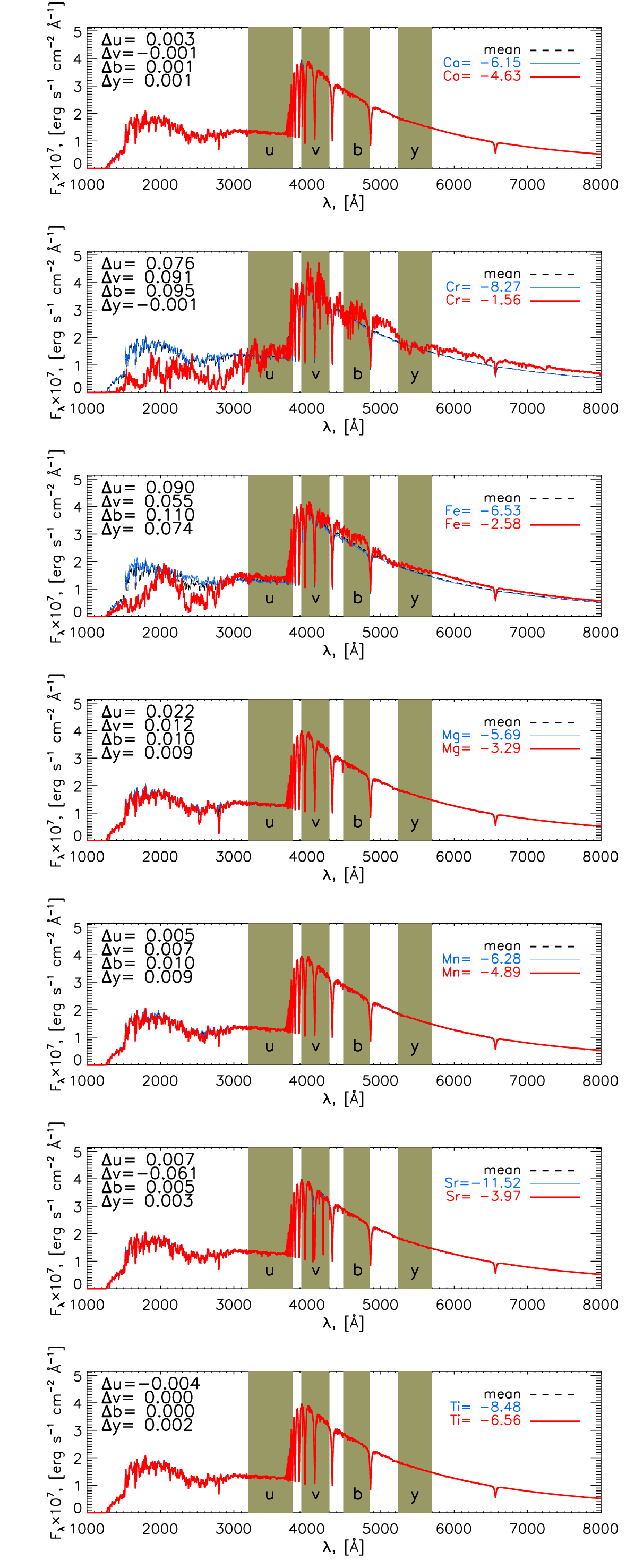}
\caption{Effect of individual elements on the synthetic energy distribution. 
The difference in fluxes in Str\"omgren $uvby$
passbands between models computed with mean abundances and maximum abundance of the particular is presented on each plot. 
Fluxes were convolved with $FWHM=20$\AA\ Gaussian for better view. The passbands (corresponding to the width of individual filter at half-maximum) 
of $uvby$ filters are indicated by grayed boxes.}
\label{fig:seds}
\end{figure}
Before presenting the results of the light curve modelling, we attempt to verify model predictions about the impact of individual
elements on overall energy distribution. Figure~\ref{fig:seds} illustrates the synthetic energy distributions for
models computed with maximum, minimum, and mean abundances of every element from DI maps. This exercise illustrates the maximum
possible energy redistribution effect caused by enhanced chemistry inside a spot. 
From Fig.~\ref{fig:seds}, one can note two general features. First, there are only two elements
that have (on average) the strongest effect on radiative energy balance: Fe and Cr. This is mainly because they have the largest
number of strong lines in the UV and thus effectively block radiation at those frequencies. Absorbed energy is then redistributed
at optical wavelengths. This is illustrated by the positive delta's of $uvby$ passbands shown on every plot. For Cr, the difference in
the $y$-band between models computed with mean and maximum Cr abundance is almost zero due to the strong Cr absorption features around 
$5500$\AA\ that keep fluxes almost unchanged for the highest abundance of $[\rm Cr]=-1.56$~dex. This never happens for
Fe, which has a  smooth flux excess in $uvby$ bands. Both elements contribute significantly to the
$5200$\AA\ flux depression frequently found in CP stars \citep[see][ and references therein]{kupka-da-2003}.

Models with enhanced Mg, Mn, and Sr cause only marginal or small changes in photometric filters. Interestingly and in contrast to
any other element, overabundant Sr leads to a noticeable flux deficiency in the $v$ parameter because of some very strong
\ion{Sr}{ii} $\lambda4078, 4162, 4216$\AA\ lines, while $uby$ passbands exhibit a far lower sensitivity to abundance changes (the same applies to the  $v$ filter in 
the Ca enhanced model because of the \ion{Ca}{ii} H \& K lines, but with much smaller amplitude). This effect is even comparable to that of
enhanced Fe. However, the mean abundance of Sr is only $-9.6$ and, on average, it is difficult to expect any noticeable impact of it on the resulting
light curve. This is investigated in more detail later in this work. Finally, the contributions of Ca and Ti to the magnitudes of
Str\"omgren photometry are very small.

Although the relative roles of different elements can be estimated 
by the aforementioned simulations, to obtain quantitative results
it is still necessary to carry out accurate modelling of light curves by taking into
account the individual contributions of all spotted elements.

\subsection{Predicted light variation}
Using DI maps of individual elements, we verified the relative contribution of each of them to the total
light variation. The result is presented in Fig.~\ref{fig:lc-ind}. 
We note that WIRE data were smoothed over $200$
points to provide more or less a reasonable view, yet some instrumental effects are still visible. 

As expected from the previous paragraph, among the seven mapped elements
there are only two that strongly contribute to the amplitude of light curves in all presented photometric
parameters: Fe and Cr. Next follows Mn, whose contribution can only be recognized by a keen eye 
(mostly in $b$ and $V$ filters). The impact of the remaining elements is negligible.
\begin{figure*}
\includegraphics[width=\hsize]{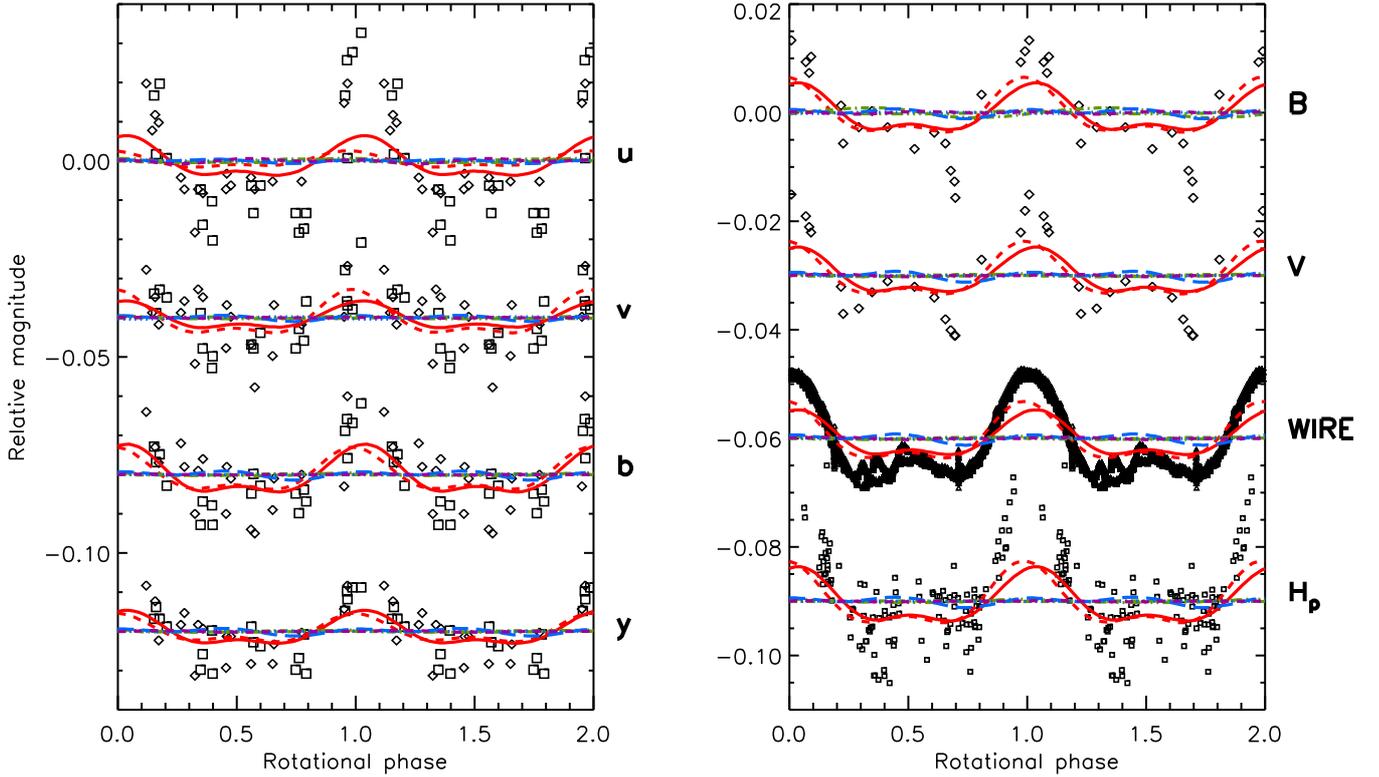}
\caption{Differential effect of inhomogeneous surface distributions of individual elements on light variations of \uma.
Fe~--~full thick, Cr~--~dashed, Ca~--~dash-dotted, Mg~--~dash-dot-dotted, Mn~--~long dashed, Sr~--~dotted, Ti~--~thin dashed.
\textbf{Left panel:}~$\Box$~--~\citet{pyper}, $\Diamond$~--~\citet{musielok1980}. 
\textbf{Right panel:}~$\Diamond$~--~\citet{provin1953}, $\triangle$~--~\citet{wire}, $\Box$~--~\citet{hip-tycho}.
\textit{See online version for color figures.}}
\label{fig:lc-ind}
\end{figure*}
The cumulative impact of all seven elements on the light curves of \uma\ is presented in Fig.~\ref{fig:lc} (thick solid line).
A very good agreement in terms of both the shape and amplitude of the light variability is found for almost  all photometric bands
(see Fig.~\ref{fig:obs.ampl}).
The only exceptions are
passbands shortward of the Balmer jump, namely
Str\"omgren's $u$, Shemakha $U_{10},$ and $P_{10}$} for which the
observed amplitude is approximately two times higher than the
predicted one (see Fig.\,\ref{fig:obs.ampl}). 
In principle, this may be a signature of some other elements that were not mapped
in \citet{lueftinger2003}, but still contribute to the light curve. Interestingly, a poorer fit to the $u$ passband
than to others was also reported by \citet{lc-hr7224} (see their Fig.~10). Nevertheless, the amplitudes in other passbands
are well reproduced in both the present work and those cited above. Taking into account the large scatter in the observed points 
 in general for the $ubvy$ system, it is difficult to explain the observed discrepancy in $u$-band in terms of impact of some missing element,
at least without additional DI mapping, if ever possible.
Finally, Hipparcos
photometry exhibits significant scatter with a peak amplitude that is a little larger than the one theoretically
predicted.

The forms of the simulated LCs agree with the observed ones very
well, all of the modelled curves being systematically shifted with respect
to the observed light curves by $+0\fp 014$.

\begin{figure*}
\includegraphics[width=\hsize]{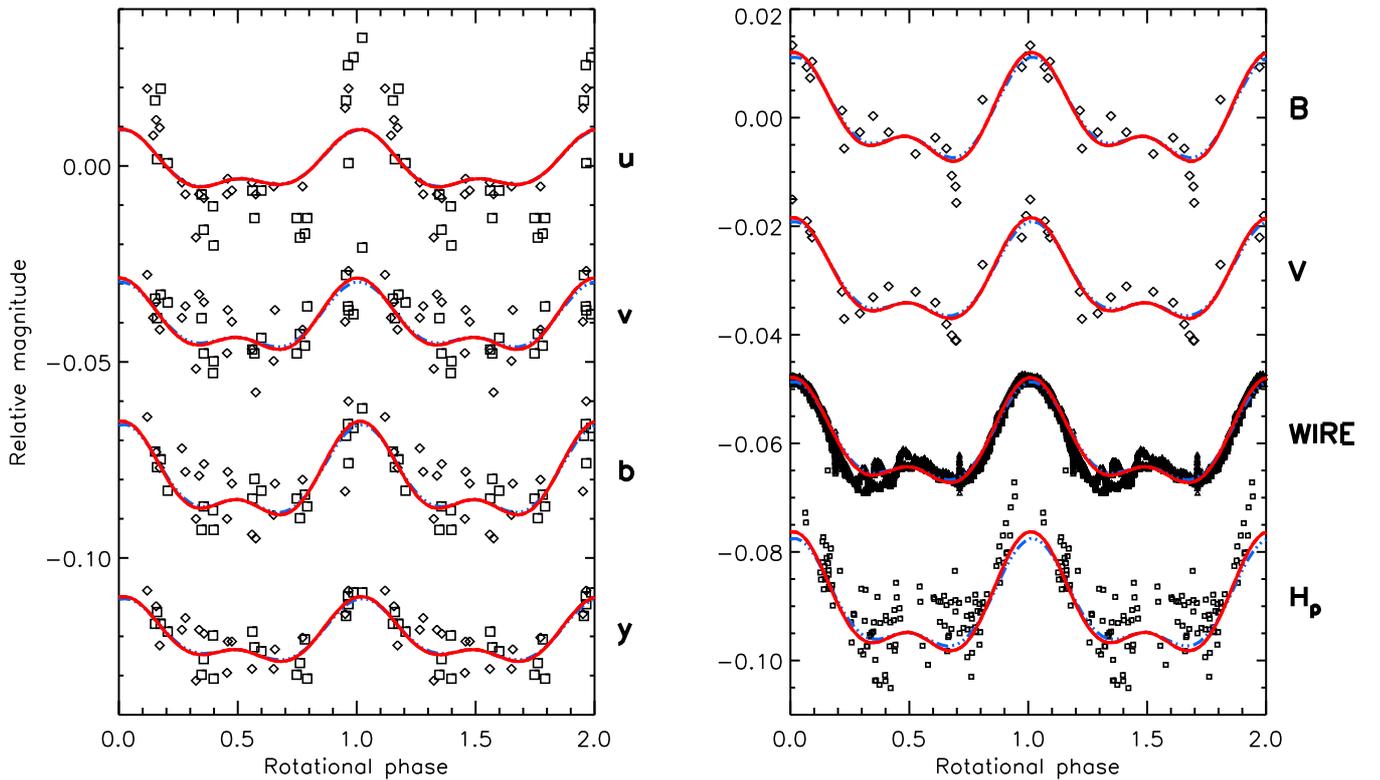}
\caption{Same as in Fig.~\ref{fig:lc-ind}, but taking into account
surface distributions of all elements simultaneously. Theoretical predictions are shown for linear (dash-dot-dotted)
and quadratic limb darkening laws based on the assumptions of [M/H]=$+0.3$ (dashed), [M/H]=$0.0$ (full thick), and
[M/H]=$-0.3$ (dash-dotted).}
\label{fig:lc}
\end{figure*}

Figure~\ref{fig:lc} allows us to conclude that the inhomogeneous surface distribution of chemical elements, being derived spectroscopically
with the help of DI techniques, can quantitatively describe the observed light variability detected in a number of 
independent photometric systems.

\subsection{UV variations}
As can be seen from Fig.~\ref{fig:seds}, the main source of light variability is
the energy redistribution from UV to visual due to the enhanced abundances
of certain elements. Thus, the fluxes in these two spectral regions should display an
anti-phase behaviour. This has been experimentally already reported for some CP stars
\citep[see][for CU~Vir and 56~Ari, respectively]{sokolov2000,sokolov2006}. 
Light curve modelling of HD~37776 and HR~7224 has also successfully predicted this characteristic signature of stellar spots.

For \uma, we again had a problem with the lack of well-calibrated phase-resolved UV observations that could be used to
study the effect of flux redistribution. The IUE archive mostly contains spectra obtained with a small aperture. This means
that some of the stellar flux was missed during individual exposures, which is important to note in our study
of the true variations in the continuum flux. Unfortunately, the only two spectra obtained with a large aperture (LWR06920RL and SWP07944RL) 
were obtained at the same rotational phase ($\varphi=0.166$).

Another possibility would be to use spectroscopic and photometric data obtained by the Copernicus OAO-2 satellite. 
This mission was equipped with a number of photometers in the UV and visual. The description of the photometric passbands
can be found in \citet{oao2-bands}. For instance, when analysing the spectroscopic data of OAO-2 in the spectral region
$1050-1830$\AA, \citet{molnar1975} detected significant flux variation caused by changes in line-blocking
as a function of the rotational phase. An anti-phase variation in particular was reported for both spectroscopic 
($1590-1690$\AA) and photometric (OAO-2 $\lambda1910$\AA\ filter) observations relative to the optical data
of \citet{provin1953}. Unfortunately, the available online co-added spectroscopic data from OAO-2 spectrometers contains
spectra obtained at different wavelength and, even worse, at different phases. We were thus unable to choose a 
unique spectral region that had been observed at a sufficiently large number of rotational phases (the case would be
a narrow region around $1200$\AA, but only $6$ phases are available and with large errors). Nevertheless,
to confirm the anti-phase behavior
of the UV flux relative to the visual, we performed an exercise similar to those presented in Fig.~5 of \citet{molnar1975}, which illustrates
the light curve of the OAO-2 $\lambda1910$\AA\ photometer. The predicted light curve was computed based on model fluxes 
in the region $1770$\AA~$\leq~\lambda~\leq~2050$\AA, which corresponds to the $FWHM$ of the OAO-2 S3F1 filter \citep[see][]{oao2-bands}.
The theoretical magnitudes are shown in Fig.~\ref{fig:oao}. The $V$-filter observations by \citet{provin1953} are also plotted.
The opposite variations in the optical and UV fluxes with respect to each other can be clearly seen. The double wave behavior of the simulated
OAO-2 curve is in excellent agreement with those presented in \citet{molnar1975}, yet the theoretical amplitude
is smaller by a factor of two. We emphasize that there is a systematical deviation between the predicted and observed magnitudes
of the OAO-2 photometry, as illustrated in Fig.~\ref{fig:obs.ampl}. Theoretical computations display a large scatter blueward of
$\log\lambda=2.4$. At the same time, some predicted points do follow the observed linear trend of the dependence of amplitude on wavelength in UV region.
This is the case for filters at $\log\lambda=2.23, 2.4, 2.5$. We note, however, that we did not attempt a detailed quantitative comparison between the observed and modelled
light curves or their amplitudes (also because the original OAO-2 filter data, as well as the filter values, are not available online), 
but only confirmed the anti-phase variations in UV and optical fluxes from the modelling point of view.

\begin{figure}[ht!]
\includegraphics[width=\hsize]{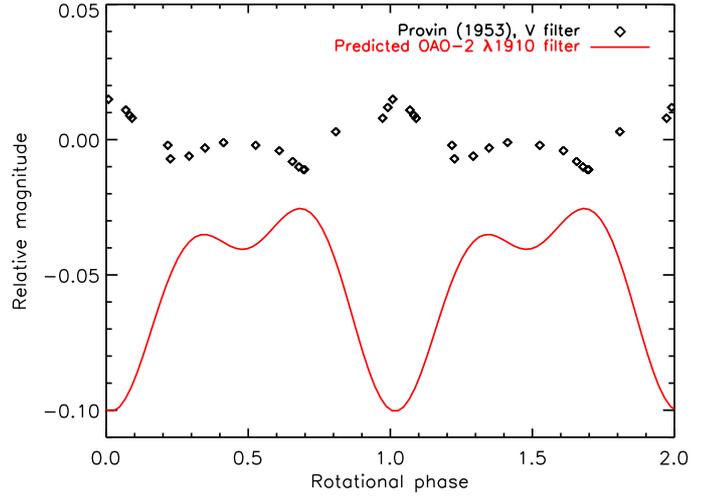}
\caption{Simulated light curves in the region of Copernicus OAO-2 $\lambda1910$\AA\ filter and optical $V$-band photometry by \citet{provin1953}.
The simulated light curve has been shifted along the y-axis for a clear visualization.}
\label{fig:oao}
\end{figure}

\subsection{Variations in the peculiar $a$-system}

The photometric parameter $a$ is often used as a measure of the peculiarity in CP stars. It was introduced
by \citet{deltaa} and corresponds to the well-known flux depression at $5200$\AA\ frequently seen in spectra
of CP stars. Using model atmosphere calculations, \citet{zeeman_paper1} noted and \citet{llind} later
confirmed (on the basis of more detailed computations) that Fe is the main contributor 
to the $5200$\AA\ depression in the temperature range of CP stars. 
In addition, at low temperatures Si and Cr are also important.
For HR~7224, \citet{lc-hr7224} also identified the dominant role of Fe relative to Si. We carried out the same investigation
and illustrate in Fig.~\ref{fig:da} the impact of different elements on the variation in the $a$-index (note that, since
the $a$-index is positive, the negative delta value in the plot correspond to the higher $a$ at corresponding phases). We thus
confirm the major role of Fe, although, the contribution of Cr is very important too. Moreover, at phase $\varphi=0.7$ the 
contribution of Cr is comparable to that of Fe.
A weak contribution of Ti can also be seen, but this fades relative to Fe and Cr. In general, the variation in the $a$-index 
appears to be small, on the order of a few mmag. A similar weak variation was reported for the hotter star HR~7224 \citep{lc-hr7224}.

\begin{figure}[ht!]
\includegraphics[width=\hsize]{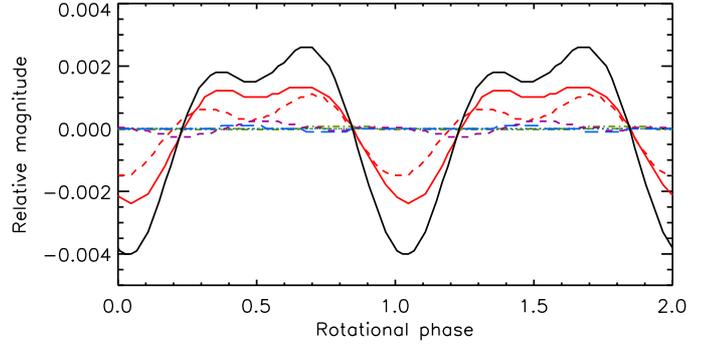}
\caption{Impact of different elements on the light curve of the peculiar $a$ parameter. 
Fe~--~full thick, Cr~--~dashed, Ca~--~dash-dotted, Mg~--~dash-dot-dotted, Mn~--~long dashed, Sr~--~dotted, Ti~--~thin dashed.
Cumulative impact of all elements is shown by the thick full thick line with highest amplitude.}
\label{fig:da}
\end{figure}

\subsection{Variation in hydrogen lines}
We found a very small variation in the $\beta$ index on the standard system defined by \citet{beta-index} related to
the Balmer \hbeta\ line,
with a peak amplitude of $\approx1.8$~mmag. This is an order of magnitude weaker than the variations
reported by \citet{musielok1988}, and may be the signature of some additional mechanism affecting
mainly profiles of hydrogen lines, as long as other photometric parameters are well fitted.
For example, the rotational modulation of hydrogen lines may be caused by
a non-zero magnetic pressure in the stellar atmosphere, as reported by \citet{lorentz-ari,lorentz-aur}.
An overview of other possible mechanisms can be found in \citet{gena2004} and we refer the
interested reader to this work for more details.
Additional observations of the $\beta$ index from \citet{musielok1980} have large errors and have
significant scatter preventing the direct comparison with \citet{musielok1988}. Last but not least,
the data from \citet{musielok1988} contains only three points at the maximum of the $\beta$ light curve
and a definite outlier at $\varphi=0.7$, which should be interpreted with caution. We thus conclude that, if true, the
high amplitude of the observed variability in the $\beta$ index cannot be attributed to abundance anomalies alone.

\section{Discussion}
\label{sec:discussion}

\uma\ is the third CP star for which a model-based light-curve analysis predicts excellent agreement
between theory and experiment, based on the assumption of the inhomogeneous surface distribution of chemical elements
derived by DI techniques. The same analysis of two other and hotter Bp stars HD~37776 \citep{lc-hd37776} and HR~7224 \citep{lc-hr7224}, 
but using slightly different methods, has also uniquely found that abundance spots may be a dominating mechanism
of light variability in CP stars.

The maximum differences between the predicted and observed light curves are seen in Str\"omgren $u$
and Hipparcos $H_{\rm P}$ bands, as illustrated in Fig.~\ref{fig:lc}. At the same time, other photometric parameters are fitted well enough
to exclude any other element significantly influencing the light curves. For instance, $BV$ parameters from \citet{provin1953}
probably provide the closest fit, again, within the error bars of the observations (see three consecutive points at
phase $1.0$, for instance). These error bars are larger for Str\"omgren passbands. \citet{lc-hr7224}
find the same larger discrepancy in $u$ but for the hotter star HR~7224. This ensures that the impact
of any missed element is less significant, yet not negligible. In principle, taking into account the relative
importance of different elements described in \citet{llind}, Si may represent such a missed element, but
it could not be mapped in \citet{lueftinger2003}. The calibration of the $u$-band may also be problematic. 
Nevertheless, the remaining $vby$ passbands can be fitted much better
without the need to involve other elements in the modelling.

As described above, to reduce computational time we decreased the resolution of the original DI maps from $2244$ surface elements
to $450$. The characteristic difference in computed magnitudes between these two sets of maps was found 
to be $\Delta c\approx 6\times10^{-4}$~mag, which is well below the detection limit.

The choice of the limb darkening model used to reconstruct the light curves in different photometric bands
can also influence the amplitude of variations. In our study, we used a quadratic law with coefficients
for every photometric band (except those described above in the paper) taken from the solar abundance model.
However, neither metallicity nor the choice of limb-darkening law seem to significantly affect the light curves, nor can they 
explain the discrepancy in $u$. This is illustrated in Fig.~\ref{fig:lc} where, as an example, we overplot the predictions 
of the linear and quadratic limb darkening laws and the solar abundance model, as well as for the quadratic law with
$[M/H]=+0.3$ and $[M/H]=-0.3$ models (note that the average metallicity derived from the DI maps is $[M/H]=+0.02$).

\section{Conclusions}
\label{sec:conclusions}

We have successfully simulated the light variability of Ap star \uma\ caused by the
inhomogeneous surface distribution of chemical elements detected by the DI maps of \citet{lueftinger2003}.
A very close agreement in both the amplitude and shape of the light variability has been found 
for different and independent photometric systems. Our main conclusions are summarized below:
\begin{itemize}
\item
The presence of abundance spots on the stellar surface is the major contributor to the light variability.
We did not introduce any free parameters to improve the agreement between theory and observations.
\item
The light variability is due to the flux redistribution from the UV to visual region, which is initiated by
the presence of abundance spots.
\item
We support the conclusion of \citet{llind} that numerous lines of Fe and Cr are the main contributors
to the well-known depression around $5200$\AA.
\item
The variation in the theoretical $\beta$ index is very weak and approximately one order of magnitude 
smaller than the observed one. We thus conclude that inhomogeneous distribution of chemical
elements alone cannot explain the observed light curve in the $\beta$ index.
\item
The strong contribution of Fe and Cr to the flux variations allows us to conclude that,
as for the hotter Bp stars that have been analysed so far (HD~37776, $\teff=22000$~K and HR~7224, $\teff=14500$~K) 
in cooler Ap stars it is also very likely that only a few elements significantly contribute to the light variation: 
Fe and Cr in the case of \uma. This must of course be tested in stars with values of $\teff$ lower than $\teff=9000$~K of \uma.
\end{itemize}

\begin{acknowledgements}
We would like to express our gratitude to Drs. Timothy Bedding and Hans Bruntt 
for kindly providing us with data from WIRE mission.
This work was supported by the following grants: Deutsche Forschungsgemeinschaft (DFG)
Research Grant RE1664/7-1 to DS, 
GAAV IAA301630901, MEB 061014 to JK and ZM.
TL thanks for support from the University of Vienna via project ``UniBRITE''.
OK is a Royal Swedish Academy of Sciences Research Fellow supported 
by grants from the Knut and Alice Wallenberg Foundation and the Swedish Research Council.
This work was supported by the financial contributions of the Austrian Agency 
for International Cooperation in Education and Research (WTZ CZ 10-2010).

We also acknowledge the use of cluster facilities at the
Institute for Astronomy of the University of Vienna, and electronic databases (VALD, SIMBAD, NASA's ADS).
Some of the data mentioned in this paper were obtained from the Multimission Archive at the Space Telescope Science Institute (MAST). 
STScI is operated by the Association of Universities for Research in Astronomy, Inc., under NASA contract NAS5-26555. 
Support for MAST for non-HST data is provided by the NASA Office of Space Science via grant NNX09AF08G 
and by other grants and contracts.
\end{acknowledgements}

%\Online
%
%--------------------------------------------------------------------
\end{document}

%% file: common-def.tex
\def\llm{{\sc LLmodels}}

\def\vald{{\sc VALD}}
\def\logg{\log g}
\def\teff{T_{\rm eff}}

\def\hbeta{H$\beta$}

%% file: tables/atmospheric-parameters.tex
\begin{table}
\caption{Stellar parameters of \uma.}
\label{tab:stellar}
\begin{footnotesize}
\begin{center}
\begin{tabular}{lc}
\hline
\hline
Effective temperature $\teff$ & $9000$~K\\
Surface gravity $\logg$ (cgs) & $3.6$\\
Inclination angle $i$         & $45^\circ$\\
\hline
\multicolumn{2}{c}{Abundance ranges from DI maps}\\
Ca & $[ -6.149,-4.628]$\\
Cr & $[ -8.270,-1.559]$\\
Fe & $[ -6.529,-2.582]$\\
Mg & $[ -5.686,-3.285]$\\
Mn & $[ -6.281,-4.890]$\\
Sr & $[-11.515,-3.967]$\\
Ti & $[ -8.478,-6.561]$\\
\hline
\end{tabular}
\end{center}
\end{footnotesize}
\end{table}

%% file: tables/phases.tex
\begin{table}
\caption{Maximum phases according to individual sources of
photometric data.
$\Delta\varphi$ is the difference between new
phases and phases evaluated according former ephemeris, see Eq.\,(\ref{eq:trafo}).}
\begin{center}
\begin{tabular}{clcrl}
  \hline\hline
  Year & max. phase & $\Delta\varphi$ & $N$ & source \\
  \hline
  1952.3 & -0.004(10)& 0.006 & 36 & \citet{provin1953} \\
  1971.0 & -0.000(3) & 0.024 & 38 & \citet{molnar1975}\\
  1975.7 & -0.001(5) & 0.028 &179 & \citet{musielok1980} \\
  1982.1 & \ 0.012(7) & 0.035 & 39 & \citet{pyper} \\
  1991.4 & -0.006(6) & 0.044 &175 & \citet{hip-tycho} \\
  1995.2 &  & 0.047 & & \citet{lueftinger2003} \\
  \hline
\end{tabular}\label{tab:OC}
\end{center}
\end{table}

%% file: tables/aeff.tex
\begin{table}
\caption{Dependence of the effective amplitude of light
variations on wavelength. $N$ denotes the number of measurements.}
\begin{center}
\begin{tabular}{crrl}
  \hline\hline
  $\lambda_{\mathrm{eff}}$ [nm]& $A_{\mathrm{eff}}$ [mag] & $N$ & effective filter \\
  \hline
  191 & 0.1232(25) & 38 & Molnar \\
  350 & -0.0336(22) & 38 & $u,\,U_{10}$\\
  360 & -0.0325(33) & 18 & $P_{10}$ \\
  410 & -0.0147(27) & 39 & $v,\,X_{10}$ \\
  440 & -0.0190(22) & 175 & $B,\,B_{\rm{T}}$ \\
  465 & -0.0218(18) & 38 & $b,\,Y_{10}$ \\
  510 & -0.0220(15) & 140 & $H_p,\,Z_{10}$ \\
  550 & -0.0209(01) & 407 & $V,\,y,\,V_{\rm{T}},\,V_{10},\,\mathit{WIRE}$ \\
  635 & -0.0230(33) & 37 & $\mathit{HR}_{10},\,S_{10}$ \\
  740 & -0.0205(28) & 34 & $\mathit{MR}_{10},\,\mathit{DR}_{10}$ \\
  \hline
\end{tabular}\label{tab:ampl}
\end{center}
\end{table}

%% file: epsilon-uHa.bbl
\begin{thebibliography}{}
\bibitem[ESA, 1997]{hip-tycho}ESA 1997, The Hipparcos and Tycho Catalogues, ESA SP-1200
\bibitem[Bessel, 2000]{hip}{{Bessell}, M.~S.}, 2000, \pasp, 112, 961
\bibitem[Bohlender \& Landstreet, 1990]{bohlender1990}Bohlender, D., Landstreet, J.~D. 1990, \apjs, 42, 421
\bibitem[Borra \& Lanstreet(1980)]{borra1980}Borra, E. F. \& Landstreet, J. D. 1980, \apjs, 42, 421
\bibitem[Claret, 2000]{claret2000}Claret, A. 2000, \aap, 363, 1081
\bibitem[Code et al., 1970]{code1970}Code, A. D., Houck, T. E., McNall, J. F., Bless, R. C. \& Lillie, C. F. 1970, \apj, 161, 377
\bibitem[Crawford \& Mander, 1966]{beta-index}Crawford, D.~L., Mander, J. 1966, \aj, 71, 114
\bibitem[Deutsch, 1947]{deutsch1947}Deutsch, A. J. 1947, \apj, 205, 283
\bibitem[Donati \& Semel, 1990]{donati1990}Donati, J.-F., Semel, M. 1990, Sol. Phys., 128, 227
\bibitem[Guthnick, 1931]{guth1931}Guthnick, P. 1931, Sitz.-Ber. Preuss. Akad. Wiss. Berlin, 27
\bibitem[Guthnick, 1934]{guth1934}Guthnick, P. 1934, Sitz.-Ber. Preuss. Akad. Wiss. Berlin, 30
\bibitem[Jamar, 1977]{jamar1977}Jamar, C. 1977, \aap, 56, 413
\bibitem[Khan \& Shulyak, 2007]{llind}Khan, S., \& Shulyak, D. 2007, \aap, 469, 1083
\bibitem[Khokhlova et al., 2000]{khokhlova2000}Khokhlova, V.L., Vasilchenko D.V., Stepanov, V.V., \& Romanyuk, I.I. 2000, AstL, 26, 177
%\bibitem[Kochukhov et al., 2004]{ok2004}Kochukhov, O., Bagnulo, S., Wade, G. A. et al. 2004a, \aap, 414, 613
\bibitem[Kochukhov et al., 2004]{ok2004}Kochukhov, O., {Drake}, N.~A., {Piskunov}, N., \& {de la Reza}, R. 2004, \aap, 424, 935
\bibitem[Kochukhov et al., 2005]{zeeman_paper1}Kochukhov, O., Khan, S., \& Shulyak, D. 2005, \aap, 433, 671
\bibitem[Kochukhov et al., 2006]{vip}Kochukhov, O., {Tsymbal}, V., {Ryabchikova}, T., {Makaganyk}, V., \& {Bagnulo}, S. 2006, \aap, 460, 831
\bibitem[{Krti{\v c}ka} et al., 2007]{lc-hd37776}{{Krti{\v c}ka}, J., {Mikul{\'a}{\v s}ek}, Z., {Zverko}, J.,	{{\v Z}i{\v z}{\v n}ovsk{\'y}}, J.} 2007, \aap, 470, 1089
\bibitem[{Krti{\v c}ka} et al., 2009]{lc-hr7224}{{Krti{\v c}ka}, J., {Mikul{\'a}{\v s}ek}, Z., {Henry}, G.~W., et al.} 2009, \aap, 499, 567
\bibitem[Kupka et al., 1999]{vald2}Kupka, F., Piskunov, N., Ryabchikova, T. A., Stempels, H. C., \& Weiss, W. W. 1999, \aaps, 138, 119
\bibitem[Kupka et al., 2003]{kupka-da-2003}{{Kupka}, F., {Paunzen}, E., {Maitzen}, H.~M.} 2003, \mnras, 341, 849
\bibitem[Kurucz, 1993]{a9-2}Kurucz, R. L. 1993, Kurucz CD-ROM 13, Cambridge, SAO
\bibitem[Leckrone, 1974]{leckrone1974}Leckrone, D. 1974, \apj, 190, 319
\bibitem[Lehmann et al., 2007]{lehmann2007}{{Lehmann}, H., {Tkachenko}, A., {Fraga}, L., {Tsymbal}, V., {Mkrtichian}, D.~E.} 2007, \aap, 471, 941
\bibitem[L\"uftinger et al., 2003]{lueftinger2003} {{L\"uftinger}, T., {Kuschnig}, R., {Piskunov}, N.~E.,	{Weiss}, W.~W.} 2003, \aap, 406, 1033
\bibitem[Maitzen, 1976]{deltaa}Maitzen, H.~M. 1976, \aap, 51, 223
\bibitem[Mallama \& Molnar, 1977]{mallama1977}Mallama, A. D., \& Molnar, M. R. 1977, \apjs, 33, 1
\bibitem[Meade, 1999]{oao2-bands}Meade, M.~R. 1999, \aj, 118, 1073
\bibitem[Michaud, 1970]{michaud1970}Michaud, G. 1970, \apj, 160, 641
\bibitem[Mikul\'a\v{s}ek, 1985]{mikthe}Mikul\'a\v{s}ek, Z. 1985, PhD Thesis, Brno
\bibitem[Mikul\' a\v sek et al., 2007a]{mikAN}Mikul\'a\v sek, Z., Jan\'{\i}k, J. Zverko, J., et al. 2007a, Astron. Nachr., 328, 10
\bibitem[Mikul\' a\v sek et al., 2007b]{miksao}Mikul\'a\v sek, Z., Krti\v cka, J., Zverko, J., et al. 2007b, in Physics of Magnetic Stars, eds. I. I. Romanyuk \& D. O. Kudryavtsev, Special Astrophys. Obs., Nizhnij Arkhyz, 300
\bibitem[Molnar, 1973]{molnar1973}{{Molnar}, M.~R.} 1973, \apj, 179, 527
\bibitem[Molnar, 1975]{molnar1975}Molnar, M.~R. 1975, \aj, 80, 137
\bibitem[Musielok et al., 1980]{musielok1980}{{Musielok}, B., {Lange}, D., {Schoenich}, W., et al.}, 1980, Astronomische Nachrichten, 301, 71
\bibitem[Musielok \& Madej, 1988]{musielok1988}Musielok, B., Madej, J. 1988, \aap, 202, 143
\bibitem[Piskunov et al., 1995]{vald1}Piskunov, N. E., Kupka, F., Ryabchikova, T. A., Weiss, W. W., \& Jeffery, C. S. 1995, \aaps, 112, 525
\bibitem[Provin, 1953]{provin1953}{{Provin}, S.~S.}, 1953, \apj, 118, 489
\bibitem[Pyper \& Adelman, 1985]{pyper}{{Pyper}, D.~M., {Adelman}, S.~J.}, 1985, \aaps, 59, 369
\bibitem[Retter et al., 2004]{wire}Retter, A., Bedding, T.~R., Buzasi, D.~L., Kjeldsen, H., Kiss, L.~L., 2004, \apj, 601, 95
\bibitem[Ryabchikova et al., 2002]{str3}Ryabchikova, T., Piskunov, N., Kochukhov, O., et al. 2002, \aap, 384, 545
\bibitem[Ryabchikova et al., 2008]{str4}Ryabchikova, T., Kochukhov, O., \& Bagnulo S. 2008, \aap, 480, 811
\bibitem[Shulyak et al., 2004]{llm}Shulyak, D., Tsymbal, V., Ryabchikova, T., St\"utz\, Ch., \& Weiss, W. W. 2004, \aap, 428, 993
\bibitem[Shulyak et al., 2007]{lorentz-aur}Shulyak, D., Valyavin, G., Kochukhov, O., et al. 2007, \aap, 464, 1089
\bibitem[Shulyak et al., 2009]{hd24712}Shulyak, D., Ryabchikova, T., Mashonkina, L., \& Kochukhov, O. 2009, \aap, 499, 879
\bibitem[Shulyak et al., 2010]{lorentz-ari}{Shulyak}, D., {Kochukhov}, O., {Valyavin}, G., et al. 2010, \aap, 509, 28
%\bibitem[Sokolov, 2008]{sokolov2008}{{Sokolov}, N.~A.} 2008, \mnras, 385, 1
\bibitem[Sokolov, 2000]{sokolov2000}{{Sokolov}, N.~A.} 2000, \aap, 353, 707
\bibitem[Sokolov, 2006]{sokolov2006}{{Sokolov}, N.~A.} 2006, \mnras, 373, 666
\bibitem[Sokolov, 2010]{sokolov2010}Sokolov, N.A. 2010, Ap\&SS, tmp, 129 (online)
\bibitem[Srivastava, 1989]{srivastava1989}{{Srivastava}, R.~K.}, 1989, \apss, 154, 333
\bibitem[Struve \& Hiltner, 1943]{struve1943}Struve, O. \& Hiltner, W. A. 1943, \apj, 98, 225
\bibitem[Swensson, 1944]{swen1944}Swensson, J. W. 1944, \apj, 97, 226
\bibitem[Valyavin et al., 2004]{gena2004}Valyavin, G., Kochukhov, O., \& Piskunov, N. 2004, \aap, 420, 993
\bibitem[Wade et al., 2000]{wade2000}Wade, G. A., Donati, J.-F., Landstreet, J. D., \& Shorlin, S. L. S. 2000, \mnras, 313, 851
\end{thebibliography}
